\documentclass[aps,pra,preprint]{revtex4}

\usepackage{epsfig}
\usepackage{subfigure}

\begin{document}

\title{Mean field master equation for self-interacting baths: comparison with exact
spin--spin-bath dynamics }
\author{Joshua Wilkie}
\affiliation{Department of Chemistry, Simon Fraser University, Burnaby, British Columbia V5A 1S6, Canada}

\date{\today}

\begin{abstract}

A mean field approximation is employed to derive a master equation suitable for
self-interacting baths and strong system-bath coupling. Solutions of the master
equation are compared with exact solutions for a central spin interacting with a spin-bath.

\end{abstract}

\maketitle

\section{Introduction}

The mostly condensed phase environments native to proposed technologies such as molecular electronics, laser control of chemical reactions, and 
quantum computing require a reexamination of the problem of decoherence. The presence of intra-environmental
coupling, for example, requires generalization beyond the uncoupled oscillator baths commonly assumed as a 
starting point in older theories\cite{OLD}. Environmental memory effects and the generally non-perturbative nature
of condensed phase system-bath interactions complicate matters even further. 

In this manuscript we use a non-perturbative mean field approximation to derive
a non-Markovian master equation for systems interacting with coupled baths. The master
equation is parameter free and preserves positivity. We test the theory by direct comparison with exact 
numerical results for a model system incorporating both intra-bath coupling and strong system-bath coupling.
Components of the theory have been published previously\cite{JW,Wilk1,Wilk2}. The complete derivation and 
comparison with exact numerical results are presented here for the first time. 

Exact arguments of Nakajima and Zwanzig\cite{Zwan} show that master equations should be
of linear integro-differential type (e.g., see Eq. (\ref{mastera}) below). In addition,
results for differential (i.e. Markovian) master equations suggest that dissipation 
should be governed by some suitable generalization of the generators for 
completely-positive-dynamical-semigroups\cite{dsg}. On the basis of these two conditions
alone one can construct a formal master equation which preserves positivity and 
has the desired form\cite{JW}. However, this introduces an
infinite number of unknown - and potentially time-dependent - parameters. Zwanzig\cite{Zwan} derived an exact integro-differential master equation but the high computational cost of calculating the dissipation terms
precludes practical application. Hence, some level of approximation of the Zwanzig equation\cite{Zwan}, consistent
with the positivity requirement, would seem a promising approach for development of 
a practical theory. Since simplifying assumptions cannot be made regarding the nature
of the bath and since system-bath coupling may be strong, a mean field approximation for
the system-bath interaction seems appropriate. This is the approach taken here.

While the master equation derived here is uniquely adapted to systems with intra-bath coupling
other methods not discussed could conceivably be modified for the same purpose. Examples
include theories which employ mixtures of quantum, semiclassical and classical dynamics for
environmental modes ranked according to their presumed importance\cite{Stock,Kapr,Wang} and approximate
density functional methods\cite{Makri}. Among master equation approaches only the Redfield
theory\cite{Opp} is potentially applicable, but it is valid only for weak system-bath
coupling, violates positivity (although this can in some instances be corrected\cite{Opp}),
predicts an incorrect long time limit\cite{Tannor2} and has dissipation coefficients which
diverge for finite baths. This last problem prevents comparison of Redfield solutions
with exact data for our model system. Finally, uncoupled harmonic baths 
seem a prerequisite for stochastic wave equation approaches\cite{Gisin}. 

Section II of this manuscript outlines the derivation of the master equation with emphasis on the basic physical
ideas. References are provided for detailed discussions of individual points. Section III
defines a model system with strong intra-bath and system-bath coupling. A numerical
approach for obtaining exact solutions is discussed. Section IV briefly describes the method employed to
solve the integro-differential master equation. Finally, section V compares the solutions of the
mean field master equation with exact solutions for the model. 

\section{Mean field master equation}

Define a projection operator $P$ on the total (system plus bath) density $\chi(t)$ such that 
\begin{equation}
P\chi(t)=\rho(t){\cal B},
\label{Pr}
\end{equation}
where $\rho(t)$ is the system density and ${\cal B}$ is the canonical bath density. Similarly, define $Q=1-P$. Assuming $\chi(0)=\rho(0){\cal B}$, a standard derivation then gives the Nakajima--Zwanzig equation\cite{Zwan}
\begin{equation}
d\rho(t)/dt=-(i/\hbar)[\bar{H},\rho(t)]-\int_0^tdt' K(t-t')\rho(t')
\label{mastera}
\end{equation}
where $\bar{H}={\rm Tr}_b\{H{\cal B}\}$ is the canonical average of the total Hamiltonian
$H$ over the states of the bath. The memory operator $K(t)$ takes the form
\begin{equation}
K(t)={\rm Tr}_b\{LQe^{-iQLQt}QL{\cal B}\}
\label{memop}
\end{equation}
where $L=(1/\hbar)[H,\cdot ]$ is the Liouville operator. Zwanzig did not use 
projection operator (\ref{Pr}) in his original derivation\cite{Zwan}, but his results are readily generalized and this operator
has been favored in the subsequent literature\cite{Pei}. The operator $QLQ$ governs the
dynamics of system-bath interaction.

Solutions of (\ref{mastera}) can only be obtained when an explicit expression for 
the memory operator (\ref{memop}) can be given. Since we cannot make simplifying
assumptions about the bath or use perturbation theory, some sort of mean field 
approximation for (\ref{memop}) seems suitable. To approximate
(\ref{memop}) we need to understand the operator $QLQ$. Consider the following
lemmas.

{\em Lemma 1: $QLQ$ is non-Hermitian.}\\
Clearly $QLQ$ is non-Hermitian if $P$ is non-Hermitian. Consider a complete orthonormal basis
$|i,j)=u_i\otimes v_j$ of the Liouville-Hilbert space where states $u_i$ and $v_j$ span the system and bath spaces, respectively. It then follows\cite{WB} that $P$ has matrix elements
\begin{equation}
(i,j|P|k,l)=\delta_{i,k}{\rm Tr}_b\{u_j^*{\cal B}\}{\rm Tr}_b\{u_l\}
\label{ma}
\end{equation}
while the matrix elements of $P^{\dag}$ are
\begin{equation}
(i,j|P^{\dag}|k,l)=\delta_{i,k}{\rm Tr}_b\{u_j^*\}{\rm Tr}_b\{u_l{\cal B}\}.
\label{mb}
\end{equation}
Since matrix elements (\ref{ma}) and (\ref{mb}) differ it follows that $P^{\dag}\neq P$
and so $P$ is non-Hermitian.

{\em Lemma 2: The spectral density of $QLQ$ is complex.}\\
The {\em average} spectral density of $QLQ$ can be defined by
\begin{equation}
\Omega(z)=\lim_{\eta\rightarrow 0}\frac{1}{\pi}\frac{\partial}{\partial
  z^*}G_{21}(z)
\label{DENS}
\end{equation}
where $G$ is an analytic $2\times 2$ average Green's function
\[\left( \begin{array}{cc}
G_{11}& G_{12}\\
G_{21}&G_{22}\end{array} \right)=\left( \begin{array}{cc}
\langle \eta [\eta^2+(z-i{\cal A})(z^*+i{\cal A}^{\dag})]^{-1}\rangle& \langle (z-i{\cal A}) [\eta^2+(z^*+i{\cal A}^{\dag})(z-i{\cal A})]^{-1}\rangle\\
\langle (z^*+i{\cal A}^{\dag}) [\eta^2+(z-i{\cal A})(z^*+i{\cal A}^{\dag})]^{-1}\rangle &\langle -\eta [\eta^2+(z^*+i{\cal A}^{\dag})(z-i{\cal A})]^{-1}\rangle\end{array} \right).
\]
Here $\eta$ is some real parameter and ${\cal A}=QLQ$. [If $z=x+iy$ then $x$ and $y$ denote imaginary and real parts of the eigenvalue of $QLQ$.] The angle brackets denote an average over the Liouville-Hilbert space i.e., for any $F$,
\begin{equation}
\langle F \rangle=\lim_{m,n\rightarrow \infty}(1/mn)\sum_{i=1}^m\sum_{j=1}^n (i,j|F|i,j) 
\end{equation}
where $|i,j)$ states denote a complete set.
Defining 
\[  G^0=\left( \begin{array}{cc}
\eta & z\\
z^*&-\eta\end{array} \right)^{-1}
\]
it can be shown that $G$ satisfies the Dyson equation
\begin{equation}
G=G^0+G^0\Sigma G,
\label{Dyson}
\end{equation}
where $\Sigma$ is a self-energy which to lowest order in $G$ is
\begin{equation}
\Sigma = \langle {\cal H}\rangle +\langle {\cal H}G{\cal
  H}\rangle -\langle {\cal H}\rangle G \langle {\cal H}\rangle
  +\dots
\label{sc}
\end{equation}
with
\[  {\cal H}=\left( \begin{array}{cc}
0 & i{\cal A}\\
-i{\cal A}^{\dag}&0\end{array}\right).
\]
Solving (\ref{Dyson}) with (\ref{sc}) trucated at first order in $G$ (self-consistent Born
approximation), and using Eq. (\ref{DENS}), one can show\cite{Wilk2} that the spectral density is uniform inside an
ellipse
\begin{equation}
\frac{x^2}{[\langle {\cal A}{\cal A}^{\dag}\rangle-\langle {\cal A}{\cal A}\rangle]^2}+\frac{y^2}{[\langle {\cal A}{\cal A}^{\dag}\rangle+\langle {\cal A}{\cal A}\rangle]^2}=\frac{1}{\langle {\cal A}{\cal A}^{\dag}\rangle}
\label{bdry}
\end{equation}
and zero elsewhere. [Simplified formulas for parameters $\langle {\cal A}{\cal A}^{\dag}\rangle$ and $\langle {\cal A}{\cal A}\rangle$, suitable for computational use, are given in Appendix A.]

Thus, $QLQ$ is non-Hermitian and its spectrum is complex in general. It then
follows that for $t\geq 0$ we may write
\begin{equation}
e^{-iQLQt}=\sum_je^{-i\omega_jt}e^{-\gamma_jt} |\phi_j)(\Phi_j| 
\end{equation}
where $\omega_j$ and $\gamma_j$ are the real and imaginary parts of an eigenvalue of $QLQ$ 
and $|\phi_j)$ and $(\Phi_j|$ are the associated right and left eigenvectors. Consequently,
the memory operator (\ref{memop}) can be written as
\begin{equation}
K(t)=\sum_j e^{-i\omega_jt}e^{-\gamma_jt} {\rm Tr}_b\{LQ  |\phi_j)(\Phi_j|QL{\cal B}\}.
\label{memop2}
\end{equation}
Obviously, for a large bath a great many terms will contribute to the sum in (\ref{memop2}).
This suggests the possibility of replacing $K(t)$ by its {\em average} (in the sense of 
Lemma 2).

Replacing (\ref{memop2}) by its average, and assuming that the statistics of the eigenvalues are independent of the eigenvectors, gives
\begin{eqnarray}
K(t)&=&\langle e^{-i\omega t}e^{-\gamma t}\rangle \langle \sum_j {\rm Tr}_b\{LQ  |\phi_j)(\Phi_j|QL{\cal B}\}\rangle \\
&=&\langle \cos (\omega t)e^{-\gamma t}\rangle {\rm Tr}_b\{LQL{\cal B}\}
\label{memop3}
\end{eqnarray}
where we have used the closure relation for the eigenvectors and the fact that for each $\omega$ there is a $-\omega$ to obtain the second equality. [We elsewhere call this the statistical resonance approximation\cite{Wilk1,Wilk2}.] Defining the memory function $W(t)=\langle \cos (\omega t)e^{-\gamma t}\rangle$ it can be shown\cite{Wilk2} that the spectral density defined in Lemma 2 gives 
\begin{equation}
W(t)=[1-\frac{4}{3\pi} (pt)^1+\frac{1}{8}(pt)^2-\frac{4}{45\pi}(pt)^3+\frac{1}{48}(pt)^4 ]e^{-(q t)^2/8}
\label{W2}
\end{equation}
where 
\begin{eqnarray}
p&=&[\langle {\cal A}{\cal A}^{\dag}\rangle-\langle {\cal
  A}{\cal A}\rangle]/\sqrt{\langle {\cal A}{\cal A}^{\dag}\rangle} \\
q&=&[\langle {\cal A}{\cal A}^{\dag}\rangle+\langle {\cal A}{\cal
  A}\rangle]
/\sqrt{\langle {\cal A}{\cal A}^{\dag}\rangle}
\end{eqnarray}
are real parameters which depend on bath temperature. This memory function is positive, satisfies $0\leq W(t)\leq 1$, and typically deviates little from gaussian form.

Finally, assuming a Hamiltonian of the form $H=H_s+H_b+\sum_{\mu}S_{\mu}R_{\mu}$ where $H_s$ and $S_{\mu}$ denote system operators and $H_b$ and $R_{\mu}$ denote bath operators, this mean field type approximation for the memory operator yields a master
equation
\begin{eqnarray}
d\rho(t)/dt&=&-(i/\hbar)[H_s+\sum_{\mu}\bar{R}_{\mu}S_{\mu},\rho(t)]\nonumber \\
&-&(1/\hbar^2)\sum_{\mu,\nu}C_{\mu,\nu}\int_0^tdt' ~W(t-t')\{[\rho(t')S_{\nu},S_{\mu}]+
[S_{\nu},S_{\mu}\rho(t')]\},
\label{masterb}
\end{eqnarray}
where $\bar{R}_{\mu}={\rm Tr}_b\{R_{\mu}{\cal B}\}$ and $C_{\mu,\nu}={\rm
  Tr}_b\{(R_{\nu}-\bar{R}_{\nu})(R_{\mu}-\bar{R}_{\mu}){\cal B}\}$
denote canonical (i.e. ${\cal B}=e^{-H_b/kT}/{\rm Tr}_b\{e^{-H_b/kT}\}$) averages and variances of bath operators.

Thus, this mean field type approximation gives a master equation in which all parameters are
known and can in principle be calculated. Moreover (\ref{masterb}) can be shown to preserve positivity 
of the density matrix\cite{Wilk2}. Given that master equation (\ref{masterb}) was obtained assuming a 
large bath, we should expect it to be most accurate in the thermodynamic limit. In the next
few sections we show that sensible results are obtained even when the number of modes of the bath
is small.

\section{Spin--Spin-Bath Model}

Our model system represents two electronic states of an atomic impurity in a crystalline solid at low temperatures. Electric-dipole transitions from the excited electronic state are forbidden, but vibronic coupling with phonons of the crystal can cause decoherence and dissipation in the impurity\cite{Tess}.

The crystal is represented by a number $n_s$ of coupled phonon modes. At low temperature the phonon modes can be roughly modeled as spin-1/2 modes (i.e., $a^{\dag}a\rightarrow \sigma_z$ and $a^{\dag}+a\rightarrow \sigma_x$) with frequencies sampled from the low energy acoustic modes of the Debye spectrum. [ We set a frequency cutoff at $\omega_D=1$.] With anharmonic phonon-phonon coupling effects included, but neglecting the zero point energies of the oscillators, our model Hamiltonian takes the form
\begin{eqnarray}
H=\frac{\omega_0}{2}\sigma_{z}^{(0)}+\beta\sigma_{x}^{(0)} + \lambda_0\sigma_{x}^{(0)}\sum_{j=1}^{n_s}\sigma_{x}^{(j)}
+ \sum_{j=1}^{n_s}[\frac{\omega_j}{2}\sigma_{z}^{(j)}+\beta\sigma_{x}^{(j)}]+\frac{\lambda}{2} \sum_{i\neq j=1}^{n_s}\sigma_{x}^{(i)}\sigma_{x}^{(j)}\label{HAM2}
\end{eqnarray}
where we arbitrarily chose $\omega_0=.8288$ as the frequency of the impurity, $\beta=.01$ is the coefficient of a small anharmonic correction, and $\lambda_0=1$ and 
$\lambda$ are the subsystem-environment and intra-environmental coupling constants. Terms one, two and four of (\ref{HAM2}) represent the uncoupled modes of the subsystem (labeled 0) and environment (labeled 1 through $n_s$). The third term in (\ref{HAM2}) couples the subsystem and environment, while the last term couples the environment with itself. The sigmas represent the Pauli matrices. In our units $\hbar=1$. Note that the environmental part of this Hamiltonian is non-integrable for $\lambda\neq 0$.

We calculated the 
reduced density matrix $\rho(t)$ of the impurity via the formula
\begin{eqnarray}
\rho(t)=\left( \begin{array}{cc}
\rho_{11}(t) & \rho_{10}(t)\\
\rho_{01}(t) & \rho_{00}(t) \end{array} \right)
=\sum_{m=1}^{n_{eig}}p_m ~{\rm Tr}_{b}\{|\psi_m(t)\rangle\langle \psi_m(t)|\}\label{DENSB}
\end{eqnarray}
where
\begin{eqnarray}
p_m=\exp\{-\epsilon_m/kT\}/\sum_{l=1}^{n_{eig}}\exp\{-\epsilon_l/kT\},
\end{eqnarray} 
$\epsilon_m$ and $|m\rangle$ are the energies and eigenvectors of the isolated environment (i.e. terms 4 and 5 of Eq. (\ref{HAM2})), and $kT$ is the temperature in units of energy. The notation ${\rm Tr}_{b}\{|\psi_m(t)\rangle\langle \psi_m(t)|\}$ indicates a trace of the full density $|\psi_m(t)\rangle\langle \psi_m(t)|$ over the environmental degrees of freedom. The states $|\psi_m(t)\rangle$ are evolved via the Schr\"{o}dinger equation from initial states
\begin{eqnarray}
|\psi_m(0)\rangle=|1\rangle\otimes |m\rangle\label{istate}
\end{eqnarray}
under Hamiltonian (\ref{HAM2}). The basis of eigenstates of the $\sigma_z$ operators was used to represent all states. The states $|0\rangle$ and $|1\rangle$ represent down and up z-components of the spin, respectively. Thus, the subsystem state $|1\rangle$ in Eq. (\ref{istate}) means that the impurity is initially in its excited state.

Equations (\ref{DENSB}) and (\ref{istate}) represent an impurity
in a thermal solid which is excited by a fast laser pulse just prior to time $t=0$ and then evolves while interacting with phonons in the solid.

The calculations reported here are for $n_s=14$ bath spins. The ARPACK linear algebra software\cite{Arp} was used to calculate the lowest $n_{eig}=20$ energies and eigenvectors of the isolated environment. A temperature of $kT=.02$ was chosen such that no states with quantum number $m$ higher than $n_{eig}$ would be populated at equilibrium. The numerical solutions of the Schr\"{o}dinger ordinary differential equations for $|\psi_m(t)\rangle$ were calculated using an eighth order Runge-Kutta routine\cite{RK}. Operations of the Hamiltonian (\ref{HAM2}) on the wavevector were calculated via repeated application of Pauli matrix multiplication routines. For example 
\begin{eqnarray}
\langle j_1,\dots,j_i,\dots,j_{n_s}|\sigma_{x}^{(i)}|\psi\rangle=\langle j_1,\dots,\bar{j_i},\dots,j_{n_s}|\psi\rangle
\end{eqnarray}
for all sets of $j_l=0,1$, $l=1,\dots ,n_s$ and where $\bar{j_i}=1$ if $j_i=0$ and $\bar{j_i}=0$ if $j_i=1$. Thus, an operation of $\sigma_{x}^{(i)}$ simply rearranges the components of $|\psi\rangle$. States of the basis 
can be represented by integers $j=j_1+j_22+\dots+j_i2^{i-1}+\dots+j_{n_s}2^{n_s-1}$ and since integers are represented in binary form on a computer, the mapping $j\rightarrow j'=j_1+j_22+\dots+\bar{j_i}2^{i-1}+\dots+j_{n_s}2^{n_s-1}$ under $\sigma_{x}^{(i)}$ can be calculated very simply
using Fortran binary-operation system functions. Operations for $\sigma_{y}^{(i)}$ and 
$\sigma_{z}^{(i)}$ are also straightforward. 

We calculated four observables. The first is the subsystem entropy 
\begin{eqnarray}
S(t)&=&-{\rm Tr}\{\rho(t)\log \rho(t)\}\label{S}\\
&=&-\frac{1}{2}\{\log {\rm det}[\rho(t)]+\sqrt{1-4{\rm det}[\rho(t)]}\log \frac{1+\sqrt{1-4{\rm det}[\rho(t)]}}{1-\sqrt{1-4{\rm det}[\rho(t)]}}\},
\end{eqnarray}
where ${\rm det}[\rho(t)]=\rho_{11}(t)\rho_{00}(t)-\rho_{10}(t)\rho_{01}(t)$,
which is initially zero because the reduced density of the subsystem is initially pure. The maximum value of this entropy is $\log 2$ which corresponds to the state
\begin{eqnarray}
\rho_{11}(t)&=&\frac{1}{2}=\rho_{00}(t)\\
\rho_{10}(t)&=&0=\rho_{01}(t).
\end{eqnarray}
The entropy gives us a quantitative measure of decoherence and dissipation effects. We also calculated the expectations of the three components of the subsystem spin
\begin{eqnarray}
X(t)&=&{\rm Tr}\{\sigma_{x}^{(0)}\rho(t)\}=\rho_{10}(t)+\rho_{01}(t)\\
Y(t)&=&{\rm Tr}\{\sigma_{y}^{(0)}\rho(t)\}=i(\rho_{10}(t)-\rho_{01}(t))\\
Z(t)&=&{\rm Tr}\{\sigma_{z}^{(0)}\rho(t)\}=\rho_{11}(t)-\rho_{00}(t).
\end{eqnarray}
The $Z(t)$ component provides information about dissipation, while the $X(t)$ and $Y(t)$ components provide information about decoherence.

\section{Numerical solution of master equation}

We recently developed a numerical technique for solving integro-differential equations\cite{TU}. The accuracy of the method has been established for both generalized Langevin
equations and master equations of type (\ref{masterb}) by comparison with exact solutions\cite{TU}. Basically the method works by
converting integro-differential equations to ordinary differential equations.

We implement the method
as follows. Define a space-like time variable $u$ and a smoothed density function
\begin{eqnarray}
\chi(t,u)=f(u)\int_0^tdt'~W(t-t'+u)\rho(t'),
\label{SMO}
\end{eqnarray}
where $f(u)$ is a damping function such that $f(0)=1$. Direct substitution shows that $\rho(t)$ and $\chi(t,u)$ satisfy ordinary differential equations
\begin{eqnarray}
&&d\rho(t)/dt=-(i/\hbar)[H_s+\sum_{\mu}\bar{R}_{\mu}S_{\mu},\rho(t)]\nonumber \\
&&~~~~~~~~~~~~-(1/\hbar^2)\sum_{\mu,\nu}C_{\mu,\nu}\{[\chi(t,0)S_{\nu},S_{\mu}]+
[S_{\nu},S_{\mu}\chi(t,0)]\},\label{tu1}\\
&&d\chi(t,u)/dt=f(u)W(u)\rho(t)+\frac{\partial \chi(t,u)}{\partial u}-\frac{f'(u)}{f(u)}~\chi(t,u)\label{tu2}
\end{eqnarray}
or more specifically for the spin--spin-bath model
\begin{eqnarray}
&&d\rho(t)/dt=-i[\frac{\omega_0}{2}\sigma_z^{(0)}+\tilde{\beta}\sigma_x^{(0)},\rho(t)]-2C\{\chi(t,0)-\sigma_x^{(0)}\chi(t,0)\sigma_x^{(0)}\} \label{stu1}\\
&&d\chi(t,u)/dt=e^{-g u^2}W(u)\rho(t)+\frac{\partial \chi(t,u)}{\partial u}+2g u~\chi(t,u)\label{stu2},
\end{eqnarray}
where $\tilde{\beta}=\beta+\lambda_0\bar{\Sigma}_x$ and $C=\lambda_0^2(\overline{\Sigma^2_x} -\bar{\Sigma}_x^2)$. Here $\Sigma_x=\sum_{k=1}^{n_s}\sigma_x^{(k)}$ and the overbar denotes a canonical average with respect to bath degrees of freedom. The parameters of the memory function (\ref{W2}) were calculated using the formulas in Appendix A and the exact energies and eigenvectors of the bath Hamiltonian computed in Section III. The same data was used to calculate $C$ and $\bar{\Sigma}_x$. Following Ref. \cite{TU} a damping function $f(u)=e^{-g u^2}$ with $g=  11/[(n-l)\Delta t]^2$ was used.
The differential equations were solved by defining a grid of points $u_j=(n+l-j)\Delta t$ with $j=1,\dots, n$ and $l=int(.338n)$ where $\Delta t=.1$ is the time-step employed in the dynamics. Converged results were obtained for $n=50$ grid points. We chose $W(u)=W(|u|)$ for negative values of $u$. A discrete-variable\cite{DVR} matrix representation 
was employed to calculate the partial derivative with respect to $u$ in Eq. (\ref{stu2}). 

Finally, the ordinary differential equations (\ref{stu1}) and (\ref{stu2}) were integrated using an eighth order Runge-Kutta routine\cite{RK}. 

\section{Results}

The exact entropy $S(t)$ is plotted in Fig. 1(a) for intra-bath couplings
$\lambda=2$ (solid curve), $\lambda=4$ (long-dashed) and $\lambda=10$ (short-dashed). While the results show a high 
degree of oscillation due to the relatively small number of bath degrees
of freedom, there is a clear trend toward smaller entropy as the intra-bath
coupling is increased. For $\lambda=2$ the entropy oscillates between zero and .5 with a mean of .25, while for $\lambda=4$ and $\lambda=10$ the mean values are roughly .08 and .006 respectively.
Figure 1(b) shows the entropy predicted by the mean
field master equation for the same values of the intra-bath coupling. The
same trend toward lower entropy with higher $\lambda$ is observed. However,
because of the mean field character of the theory no oscillations are
observed. 

An explanation of this trend toward lower entropy
for larger intra-bath coupling is presented elsewhere\cite{Tess}.

\begin{figure}[htp]
\caption{$\lambda=2,4,10$}
\subfigure[$S(t)$]{
\epsfig{file=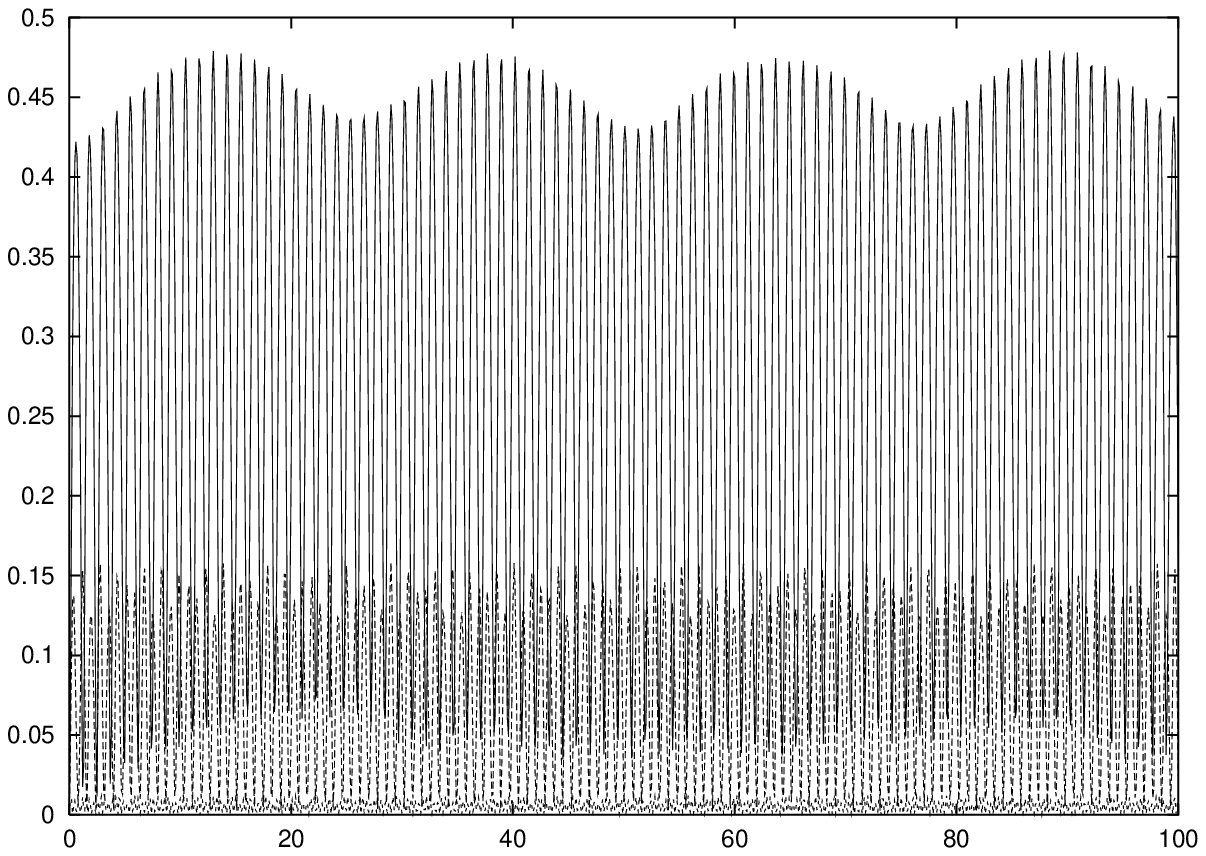,width=3.15in,height=2.5in}}
\subfigure[$S(t)$]{
\epsfig{file=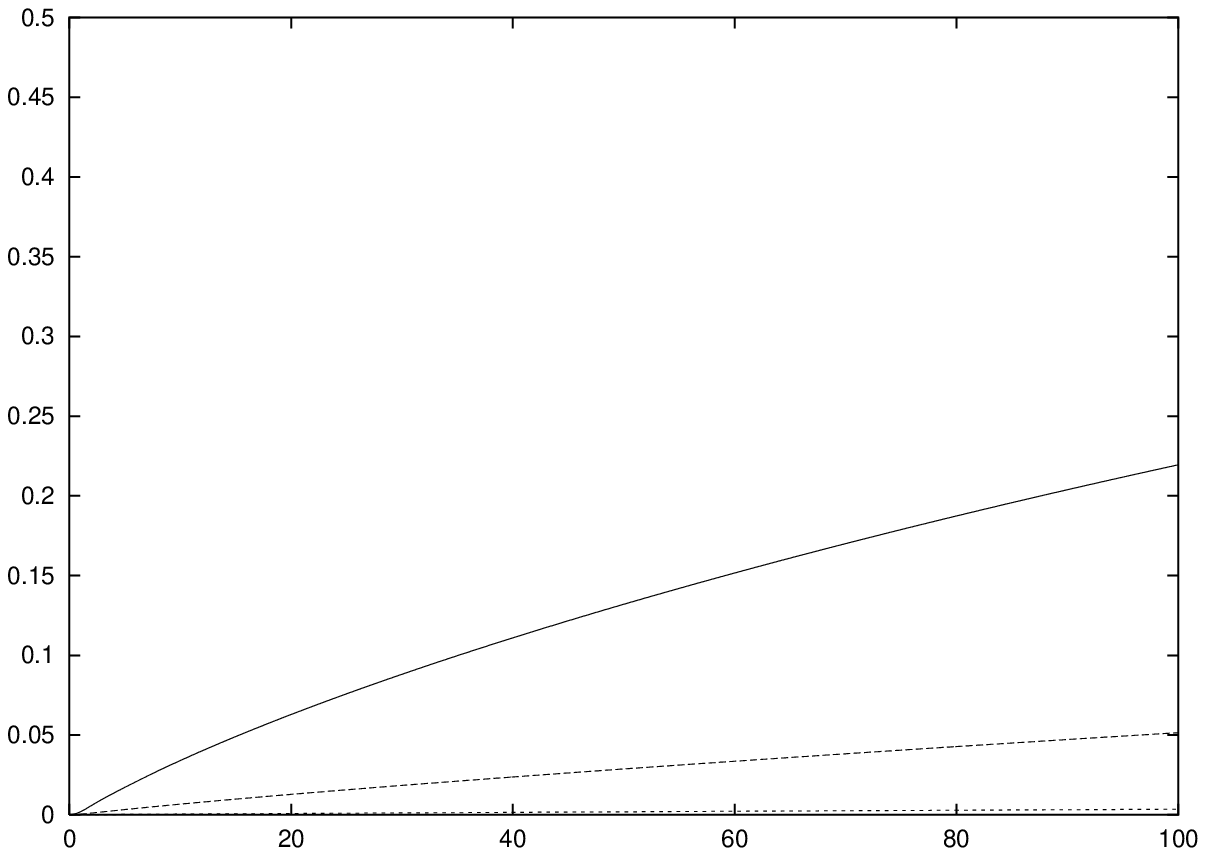,width=3.15in,height=2.5in}}
\end{figure}

Expectations of the components of the subsystem spin are plotted in Fig. 2
for $\lambda=2$ for exact (long-dashed) and mean field (short-dashed) calculations. For reference we also show the spin
dynamics in the absence of system-bath coupling (solid curve). Clearly, the exact and 
mean field results show strong decoherence of comparable magnitude. After
a short time the mean field $X(t)$ and $Y(t)$ become phase shifted from
the exact results which also show evidence of noise. Amplified oscillations
of similar character are observed in the exact $Z(t)$ but are absent in the mean
field solution. The decoherence-free solution is indistinguishable from the upper boundary
of the figure.

\begin{figure}[htp]
\caption{$\lambda=2$}
\subfigure[$X(t)$]{
\epsfig{file=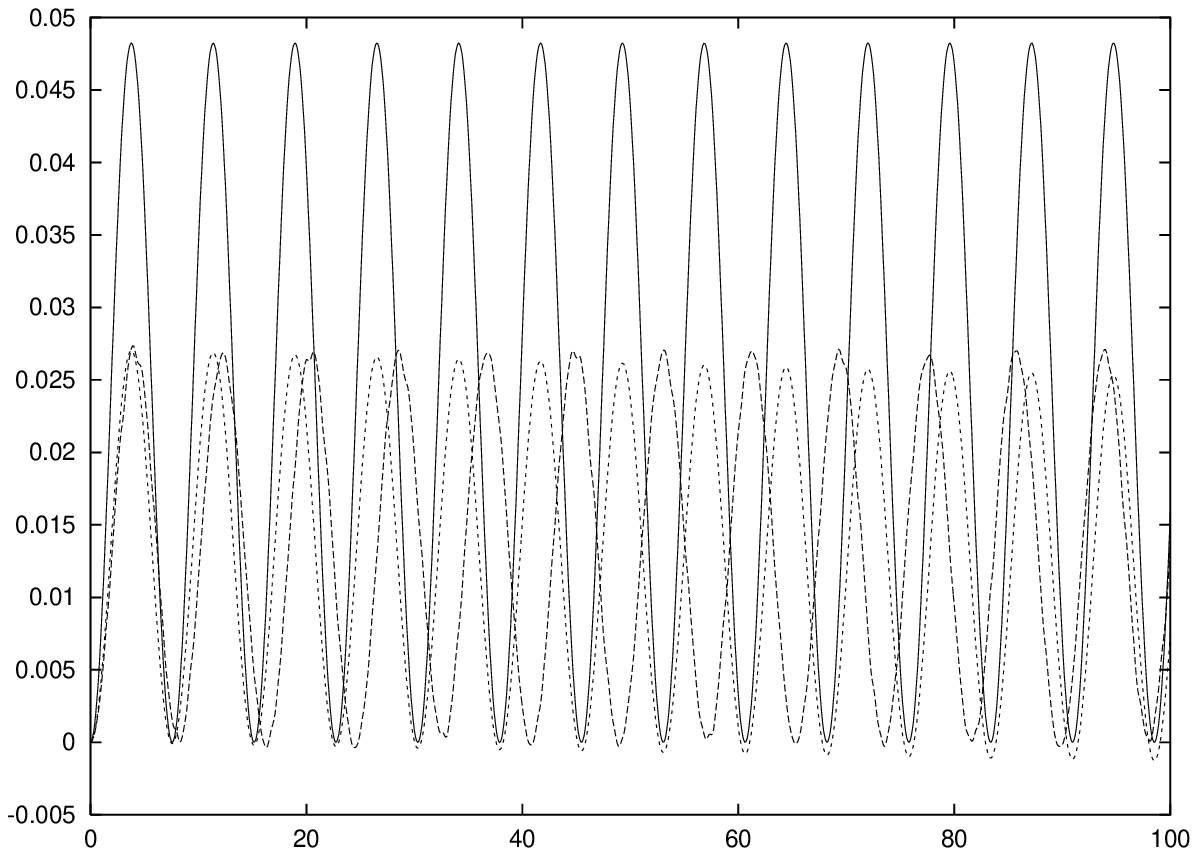,width=3.15in,height=2.5in}}
\subfigure[$Y(t)$]{
\epsfig{file=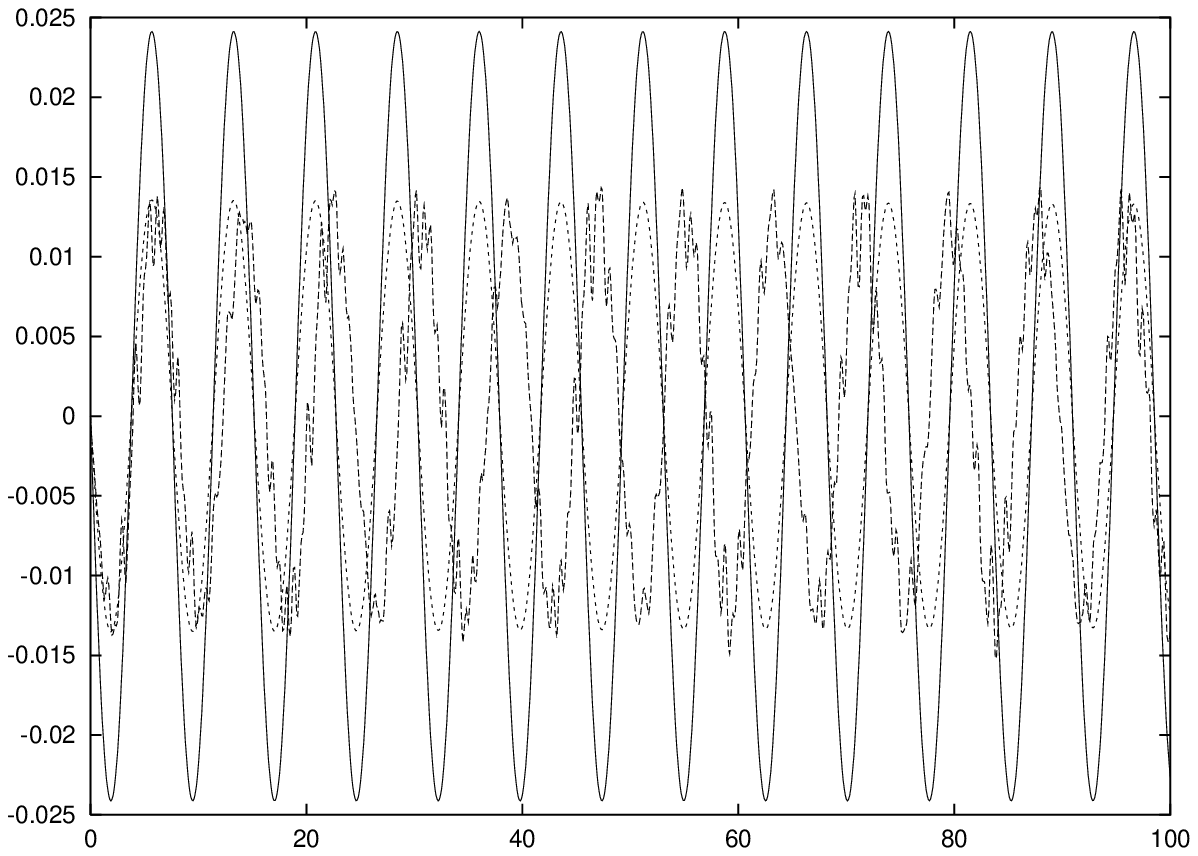,width=3.15in,height=2.5in}}
\subfigure[$Z(t)$]{
\epsfig{file=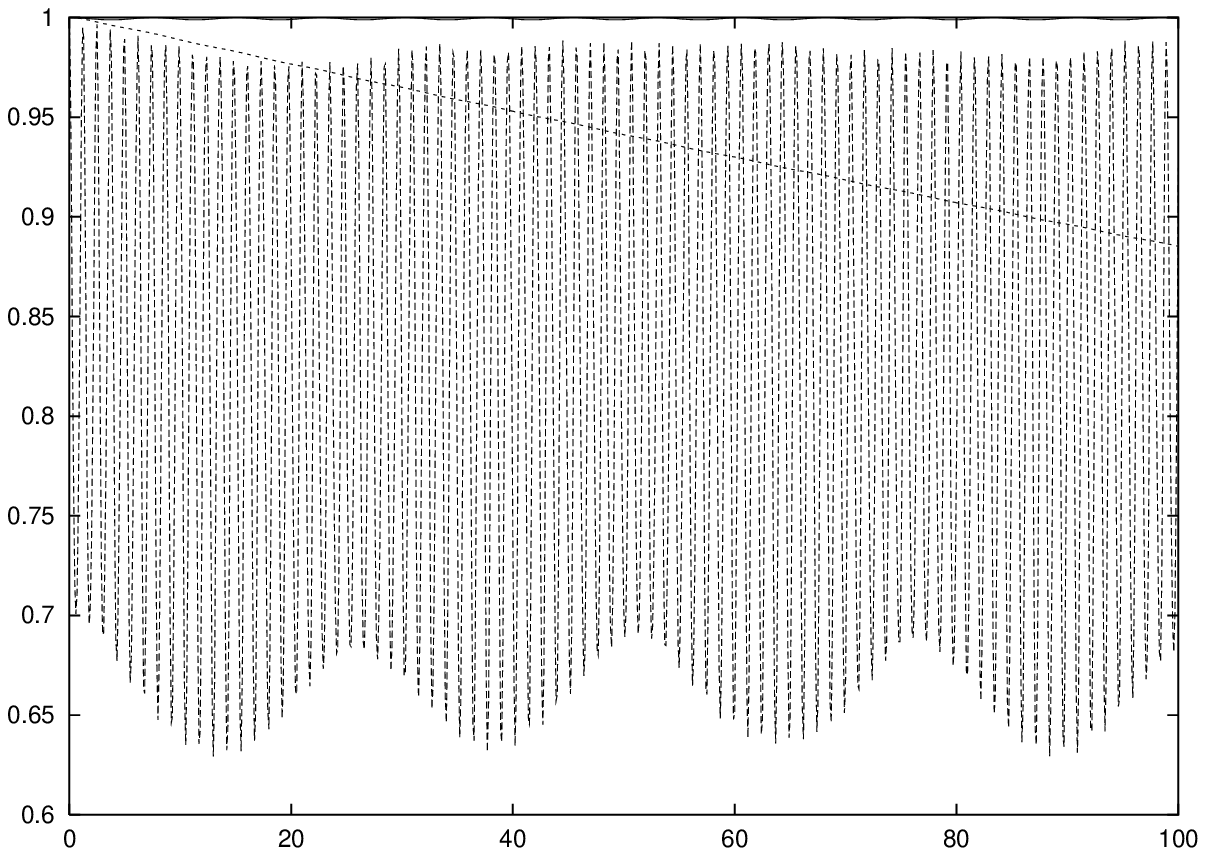,width=3.15in,height=2.5in}}
\end{figure}

Similar calculations are shown in Fig. 3 for $\lambda=4$ and in Fig. 4 for
$\lambda=10$. Decoherence of $X(t)$ and $Y(t)$ is incementally decreased
in both exact and mean field solutions which show increasingly good 
agreement. Reduced dissipation in $Z(t)$ is predicted by both calculations
as $\lambda$ increases.

\begin{figure}[htp]
\caption{$\lambda=4$}
\subfigure[$X(t)$]{
\epsfig{file=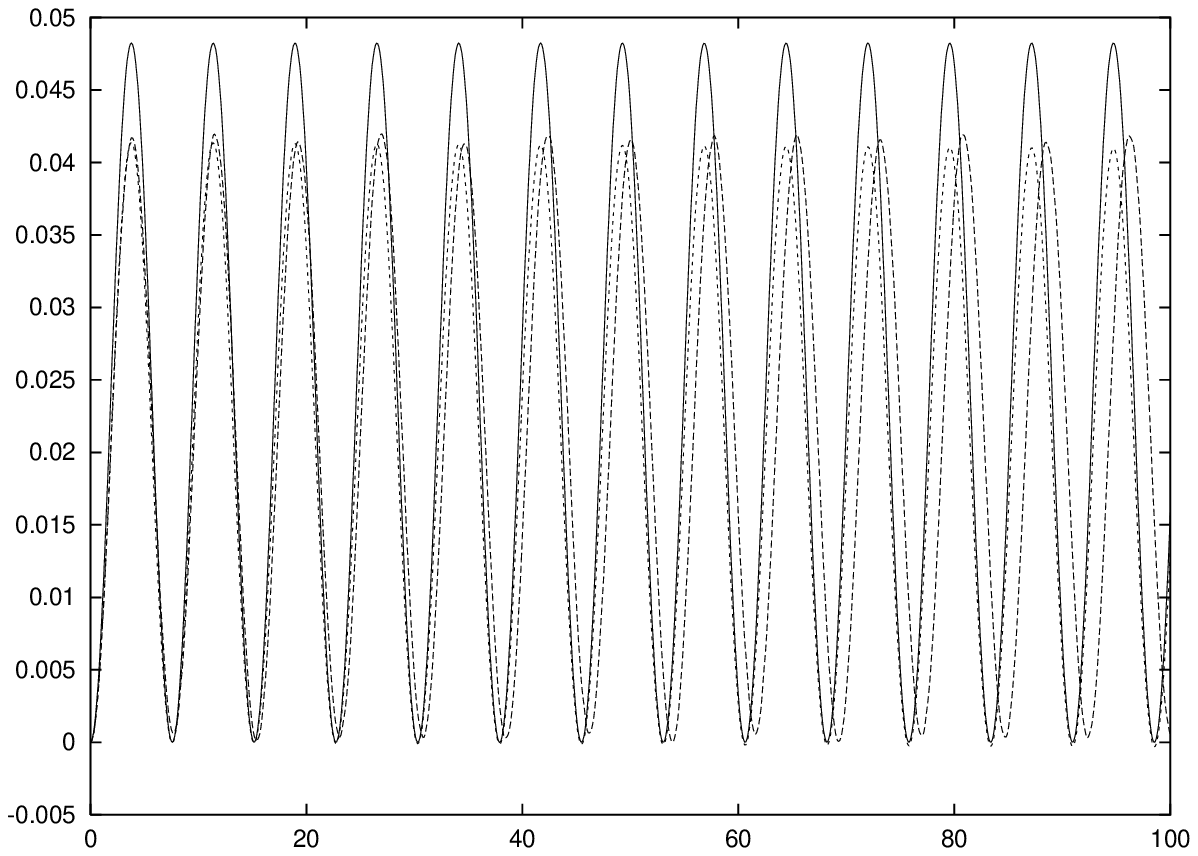,width=3.15in,height=2.5in}}
\subfigure[$Y(t)$]{
\epsfig{file=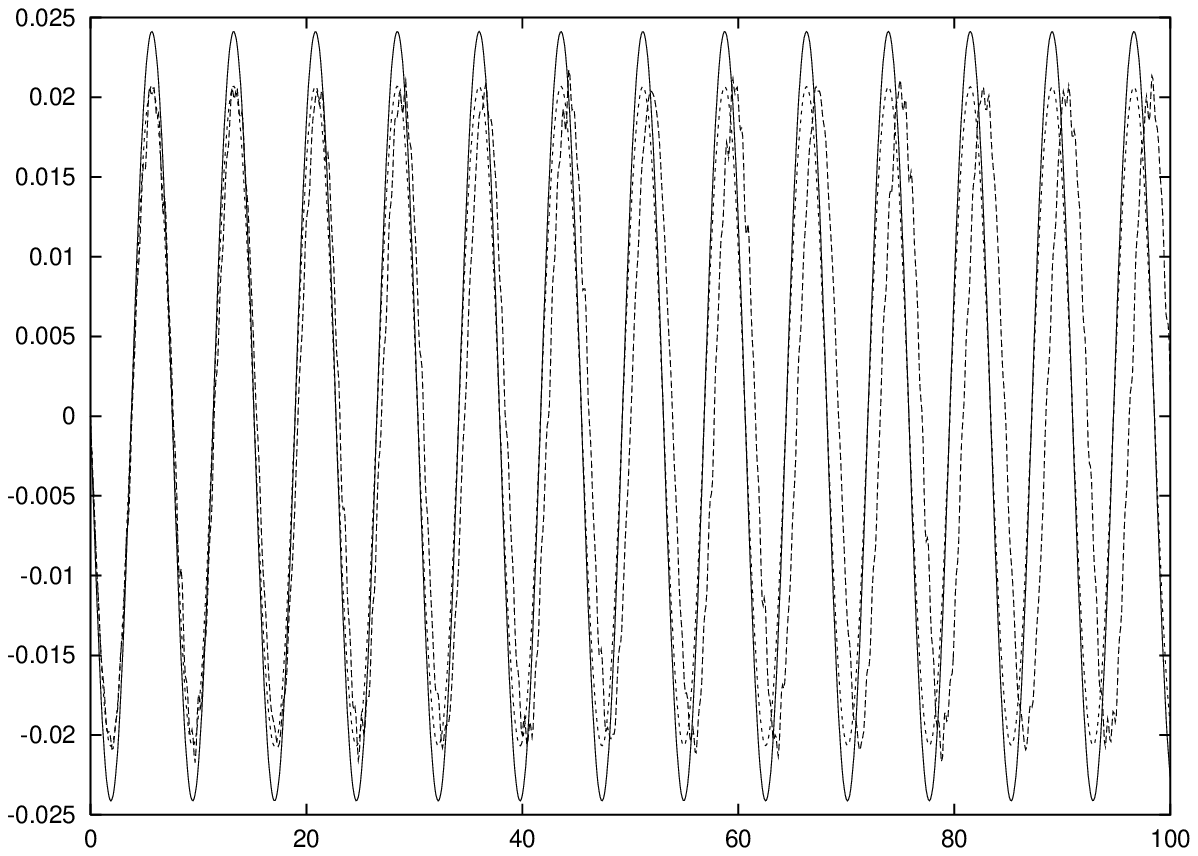,width=3.15in,height=2.5in}}
\subfigure[$Z(t)$]{
\epsfig{file=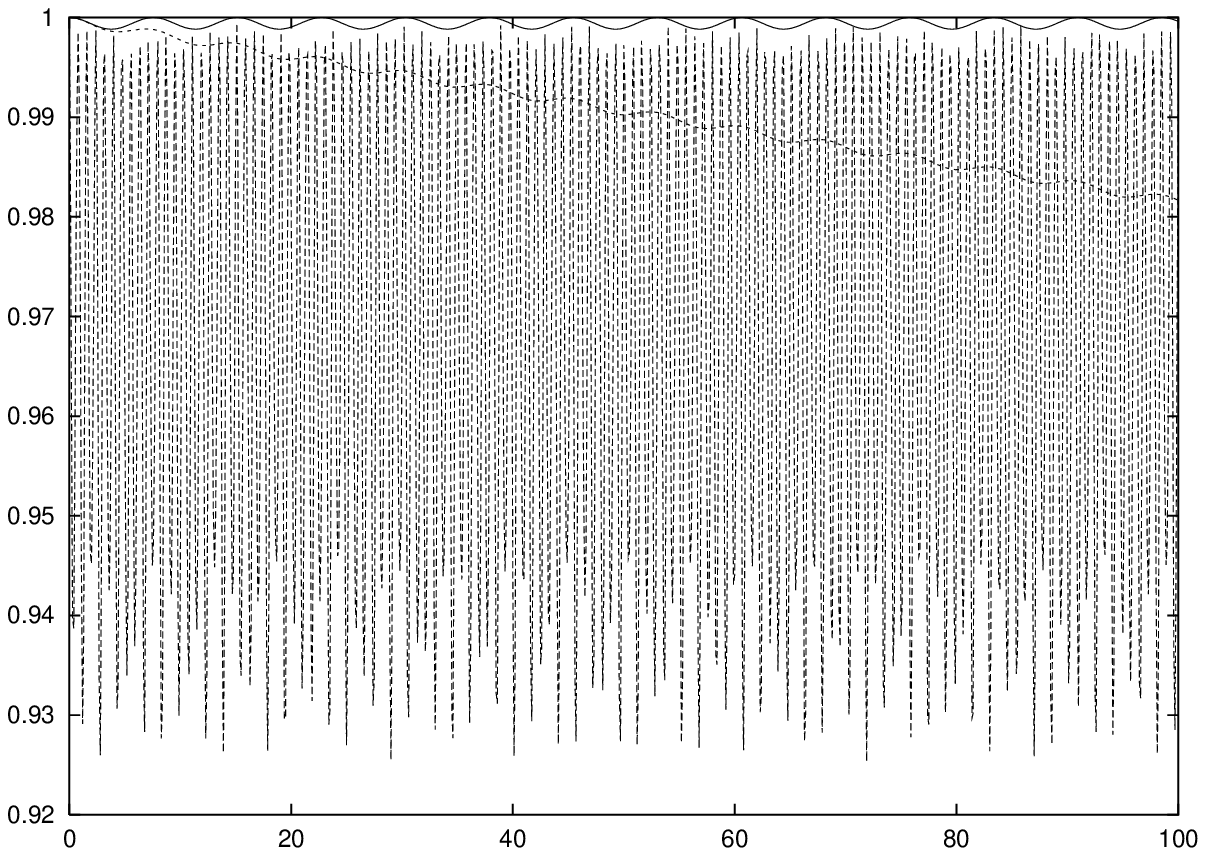,width=3.15in,height=2.5in}}
\end{figure}

Given the relatively small size of the present bath and the consequently oscillatory character of the observables, and supposing that the 
mean field approximation will improve with larger baths, these results 
appear to support the general approach of treating the system-bath
interaction in a mean field approximation.

\begin{figure}[htp]
\caption{$\lambda=10$}
\subfigure[$X(t)$]{
\epsfig{file=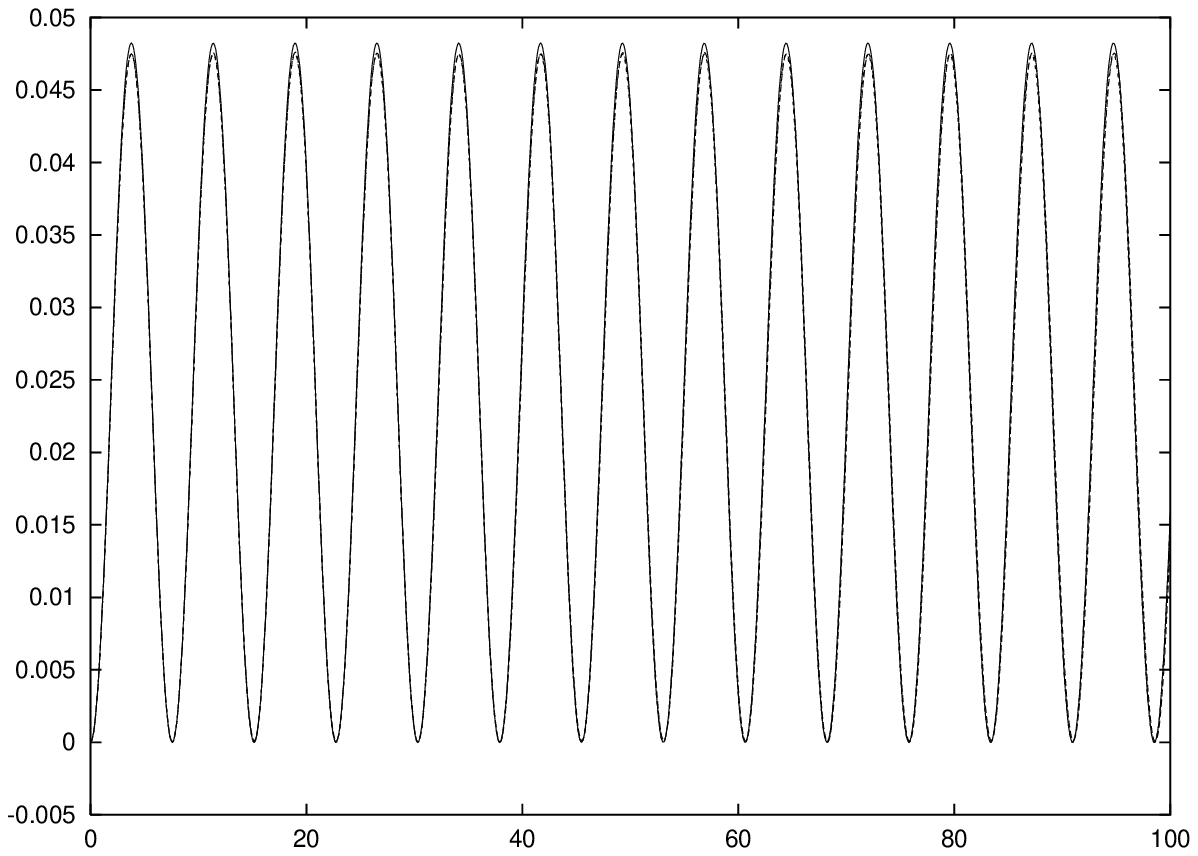,width=3.15in,height=2.5in}}
\subfigure[$Y(t)$]{
\epsfig{file=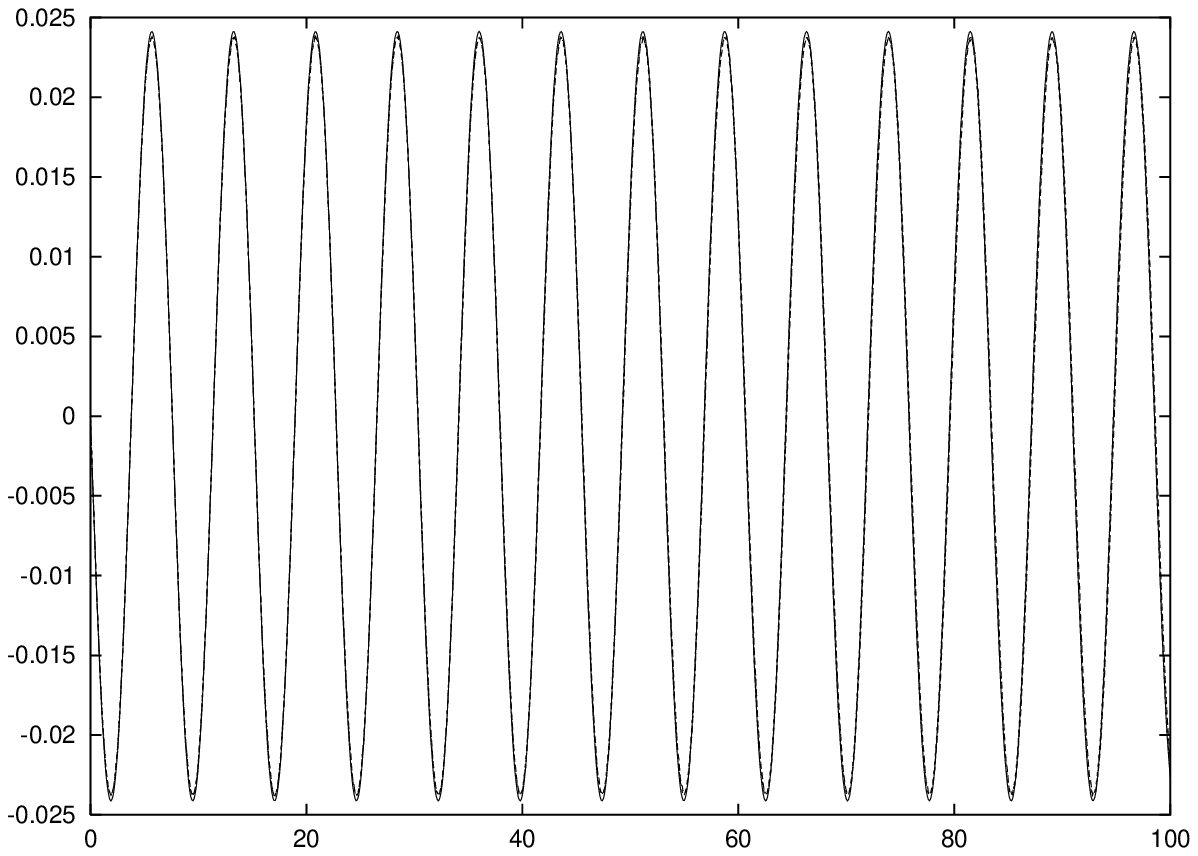,width=3.15in,height=2.5in}}
\subfigure[$Z(t)$]{
\epsfig{file=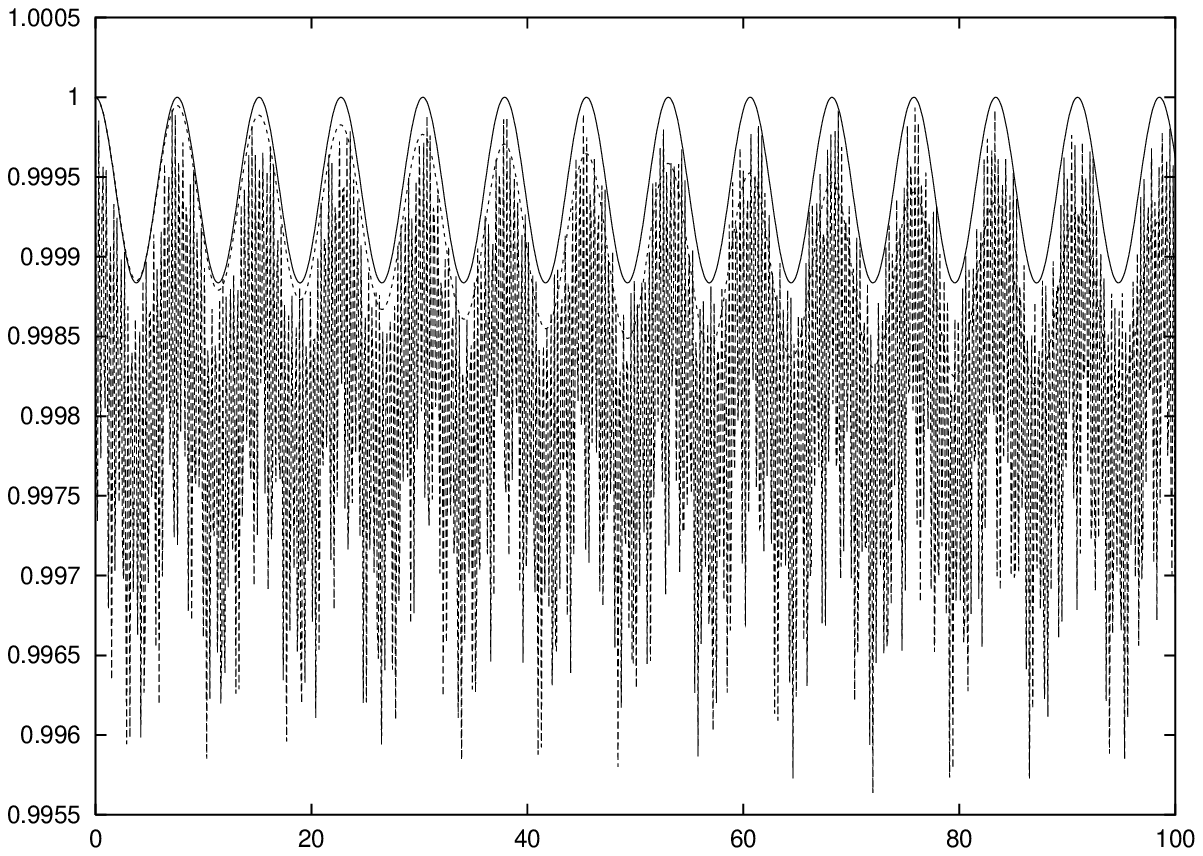,width=3.15in,height=2.5in}}
\end{figure}

\section{Summary}

The problem of predicting the dynamics of a system interacting with a 
condensed phase environment requires a reexamination of the assumptions
commonly employed in theories of decoherence and dissipation. Specifically,
intra-bath coupling cannot be neglected {\em a priori}, system-bath coupling may be strong,
and memory effects may play a role. Simplifications employed in older theories
such as modeling the bath as uncoupled oscillators, assuming the validity
of perturbation theory in the system-bath coupling, and use of Markovian
assumptions must therefore be abandoned in general. This raises the issue
of how an appropriate theory can be derived.

In this manuscript we introduce a sort of mean field approximation
in the system-bath coupling and use it to obtain an approximate master 
equation which preserves positivity. The predictions of the master
equation are tested against exact results for a model system consisting
of a spin interacting with a spin-bath. In spite of the oscillatory
character of some of the system observables good qualitative agreement
is observed, raising the possibility that the master equation may prove 
quantitatively accurate for larger baths. We hope to soon further 
test the theory against exact results for a coupled oscillator bath using
a recently developed exact method\cite{PRE} for decomposing the $N$-vibrational-mode time evolving density matrix (for pairwise interactions) into $N$ one-dimensional stochastic 
density equations.

The author gratefully acknowledges the support of the Natural Sciences and 
Engineering Research Council of Canada.

\section{Appendix A}

Explicit formulas for $\langle {\cal A}{\cal A}^{\dag}\rangle$ and $\langle {\cal A}{\cal A}\rangle$ in terms of averages over finite basis sets of the Hilbert space are as follows:
\begin{eqnarray}
\langle {\cal A}{\cal A}^{\dag}\rangle 
&=& 
\frac{1}{m_s^2m_b^2} 
[
2m_sm_b{\rm tr}\{H^2\}
-2{\rm tr}\{H\}^2
-8m_s{\rm tr}\{H^2{\bf B}\} 
-4{\rm tr}\{H{\bf B}\}^2
+2m_sm_b{\rm tr}\{H^2{\bf B}^2\}\nonumber \\
&+&
2m_s {\rm tr}\{H^2\}{\rm tr}_b\{ {\bf B}^2\}
+4m_b{\rm tr}\{H{\bf B}\}{\rm tr}\{H{\bf B}^2\}
+8{\rm tr}_b\{{\rm tr}_s\{H\}^2{\bf B}\}
+4m_s {\rm tr}_s\{ {\rm tr}_b\{ H{\bf B} \}^2 \} 
\nonumber \\
&+&
4m_s{\rm tr}_s\{{\rm tr}_b\{H{\bf B}^2\}{\rm tr}_b\{H\}\}
-4{\rm tr}\{H\}{\rm tr}\{H{\bf B}^2\}
-2m_b{\rm tr}_b\{{\rm tr}_s\{H{\bf B}\}^2\}
\nonumber \\
&-&
2{\rm tr}_b\{{\rm tr}_s\{H\}^2\}{\rm tr}_b\{{\bf B}^2\}
-4m_sm_b {\rm tr}_s\{{\rm tr}_b\{H{\bf B}\} {\rm tr}_b\{H{\bf B}^2\}\}\nonumber \\
&-&4m_s{\rm tr}_s\{{\rm tr}_b\{H\} {\rm tr}_b\{H{\bf B}\}\}{\rm tr}_b\{{\bf B}^2\}
+
2m_sm_b {\rm tr}_s\{ {\rm tr}_b\{ H {\bf B}\}^2 \} {\rm tr}_b\{ {\bf B}^2 \}\nonumber \\
&-&2m_b {\rm tr}\{ H{\bf B} \}^2 {\rm tr}_b\{ {\bf B}^2 \} +4{\rm tr}\{H\}{\rm tr}\{H{\bf B}\}{\rm tr}_b\{{\bf B}^2
\}] \label{AAD}\\
\langle {\cal A}{\cal A}\rangle 
&=&
\frac{1}{m_s^2m_b^2}[
2m_sm_b{\rm tr}\{H^2\}
-2{\rm tr}\{H\}^2+4{\rm tr}_b\{{\rm tr}_s\{H\}^2{\bf B}\}
+2m_s {\rm tr}_s\{{\rm tr}_b\{H{\bf B}\}^2\} \nonumber \\
&-&4m_s{\rm tr}\{ H^2{\bf B}\}
- 2{\rm tr}\{ H{\bf B}\}^2
].\label{AA}
\end{eqnarray}
Here the traces are over the normal Hilbert space rather than the Liouville-Hilbert space.
Specifically, the finite basis ($m_s$ basis functions) trace over subsystem degrees of freedom is denoted ${\rm tr}_s\{\cdot \}$, the finite basis ($m_b$ basis functions) trace over reservoir degrees of freedom is denoted ${\rm tr}_b\{\cdot\}$, and the complete trace over the finite basis ($m_s\times m_b$ functions) is denoted ${\rm tr}\{\cdot \}$. It is essential that all matrices be represented in the finite basis before the calculations for (\ref{AAD}) and (\ref{AA}) are carried out. The size of the finite basis should be chosen so that higher energy states are unpopulated at the given temperature.

\end{document}